\documentclass[aps,pre,reprint,groupedaddress,showpacs]{revtex4-1}

\usepackage{amsmath,amsfonts,amssymb}
\usepackage{graphicx}
\usepackage{natbib}
\usepackage{hyperref}

\begin{document}

\title{Dispersion of particles by a strong explosion}

\author{Timothy C. DuBois}

\author{Milan Jamriska}

\author{Alex Skvortsov}
\email[]{alex.skvortsov@dsto.defence.gov.au}
\affiliation{Defence Science and Technology Organisation, 506 Lorimer Street, Fishermans Bend, Vic 3207, Australia}

\date{\today}

\begin{abstract}
The dynamics of particle transport under the influence of localised high energy anomalies (explosions) is a complicated phenomena dependent on many physical parameters of both the particle and the medium it resides in.
Here we present a conceptual model that establishes simple scaling laws for particle dispersion in relation to the energy released in a blast, properties of the medium, physical properties of particles and their initial position away from a blast epicenter.
These dependencies are validated against numerical simulations and we discuss predictions of the model which can be validated experimentally.
Other applications and extensions to the framework are also considered.
\end{abstract}

\pacs{47.70.-n, 05.60.-k, 92.60.Mt}

\maketitle

\section{Introduction}

Understanding transport properties of particle systems driven by strong energy fluxes is of significant importance to a number of fields of science and technology.
Examples include non-equilibrium statistical physics \cite{Krapivsky10, Falkovich01}, astrophysical and geophysical phenomena \cite{Chapman94,Dunne03,Nozawa07}, multi-phase turbulent flows \cite{Zaichik09}, inhomogeneous catalysis \cite{Bird06}, combustion \cite{Balakrishnan10} and many others \cite{Grebenkov14,Gottiparthi14,Zukas97,Zhang01}.
An important requirement for these studies is the development of a rigorous framework which estimates system parameters (\textit{e.g.}\ energy fluxes) from remote (or retrospective) observations of particles following natural or anthropogenetic phenomena (a volcano eruption, meteorite impact, supernova event, blast, \textit{etc}).
Two revealing examples of this approach are the well-known pioneering studies of L.F.~Richardson (estimation of parameters of turbulent flows from Lagrangian measurements \cite{Richardson26}) and E.~Fermi (a remote estimation of a nuclear bomb yield from `tracer' particle observations~\cite{Archives45}).

There is a vast amount of literature devoted to the subject of particle transport (see \cite{Krapivsky10, Zaichik09, Balakrishnan10,Shukla02,Bird06} and references therein). Modern computational models (often called models of Lagrangian transport) achieve an unprecedented level of fidelity by matching numerical predictions with experimental observations \cite{Balakrishnan10,Toschi09, Zhang01}.
Unfortunately, whilst these computational models are an important predictive tool for practical applications and validation studies, they are unable to provide analytical insights into the fundamental transport mechanisms of these systems -- simply because analytical predictions cannot be deduced numerically.
Significant analytical progress in the understanding of transport phenomena in particle systems has been achieved by employing scaling and self-similarity frameworks \cite{Krapivsky10,Barenblatt97,Falkovich11}. This allows us to describe the dispersion process by means of power-law functions (scaling laws) relating to particle displacement and other parameters of the system -- the exponent of these power-laws being predicated analytically.
Using this knowledge in conjunction with Lagrangian measurements, one can infer values of important system parameters that would be challenging to recover by any other means.

The presented results are in line with this approach.
More specifically: we establish a scaling law for particle dispersion caused by a rapid (and localized) energy release (explosion), and express it as a simple power-law relation between the particle displacement and the physical parameters of the system (energy of the explosion, particle properties and their initial position, \textit{etc}).
We demonstrate that under a broad range of conditions the exponents of this scaling law can be deduced analytically by applying the ideas of self-similarity.
We support our analytical predictions with numerical simulations.

\section{Model}\label{sec:model}

The following is a simplified conceptual model that allows us to derive a scaling law for particle displacement.
We consider an infinite domain initially populated with particles whose density is much larger than the media density (we assume this media to be a gas with known properties).
We restrict ourselves to the case when the density fraction of particles is relatively small, so particle-particle interactions can be neglected and thus consider the dispersion of a single tracer particle (the opposite limit of a localized energy release in a `crowded' system of particles was analysed in Ref.~\onlinecite{Antal08}.
Similar to other studies \cite{Grafke14}, we assume that the dynamics of an inertial particle are dominated by viscous drag of the parent medium and is described by a force equation.
For the sake of simplicity we disregard all other processes that may occur in the system (\textit{e.g.}\ multi-phase transitions).

The equation of motion for a particle takes the standard form

\begin{equation}
\dot{\textbf{r}} = \textbf{v}, \quad \dot{\textbf{v}} = \frac{1}{\tau}(\textbf{V}  - \textbf{v}),
\label{eq:force}
\end{equation}
where $\tau$ is the Stokes time of the particle, $\textbf{V}(t) \equiv \textbf{V}(\textbf{r}(t), t)$ is the velocity field induced by a blast at the position of the particle.

We assume that the flow field velocity $\textbf{V}(\textbf{r}, t)$ can be approximately described by the Sedov-Taylor solution for a strong explosion~\cite{Landau87}

\begin{equation}
\textbf{V}= \frac{2\textbf{r}}{5t} \Psi(\zeta),
\label{eq:vflow}
\end{equation} 
where $\zeta = r/R(t)$ (and $r\!=\!0, \: t\!=\!0$ corresponds to the initial location and ignition time of the blast).

\begin{equation}
\label{eq:Rt}
R = \beta\left(\frac{Et^2}{\rho}\right)^{1/5},
\end{equation}
where $R(t)$ is the position of the shock front, $E$ is the total energy released in the blast and $\rho$ is the density of the medium.
The dimensionless parameter $\beta$ is function of the polytropic exponent $\gamma$ (for $\gamma = 7/5$: $\beta \approx 1.033$ \cite{Landau87}) and the function $\Psi(\zeta)$ can be closely approximated by its limiting value $\Psi(\zeta) \approx 1/\gamma$, if $0 \le \zeta \le 1$, and $\Psi(\zeta)=0$ otherwise~\cite{Landau87}.
The velocity of the shock front is given by the derivative of expression (\ref{eq:Rt}):
\begin{equation}
\label{eq:dRdt}
\dot{R} =\frac{2  \beta }{5} \left(\frac{E}{\rho t^3}\right)^{1/5}.
\end{equation}


For a particle located at an initial position $r=r_0$ away from the blast ignition point at $t=0$,  Eq.~(\ref{eq:force}) reduces to the scalar form

\begin{equation}
\ddot{r} + \frac{\dot{r}}{\tau} - \frac{q}{\tau} \frac{r}{(t + t_0)} = 0,
\label{eq:ODE}
\end{equation}
where $q = 2/(5\gamma)$ and $t_0$ is the time required for the shock wave to reach the particle, estimated from Eq.~(\ref{eq:Rt})

\begin{equation}
\label{eq:t0}
t_0 = \left(\frac{r_0}{\beta}\right)^{5/2}\left(\frac{\rho}{E}\right)^{1/2}.
\end{equation}

From Eq.~(\ref{eq:ODE}) we can readily deduce a scaling law for particle displacement when inertial effects are negligible.
By dropping the first term in this equation we arrive at

\begin{equation}
\frac{r}{r_{ref}} = \left(\frac{t}{t_{ref}}\right)^q,
\label{eq:scaleref}
\end{equation}
where $r_{ref}$ is some arbitrary reference position which defines the particle location at time $t=t_{ref}$.
The $r_{ref}$ and $t_{ref}$ scales have been introduced to satisfy two initial conditions of the original equation (\ref{eq:ODE}).
If we define $t_{ref} \equiv t_0 + \tau$ (\textit{i.e.} time when inertial effects become unimportant) then $r_{ref} = r_0 + r_\tau$, where $r_\tau$ is the particle displacement during the Stokes time $\tau$ (\textit{i.e.}\ from $t_0$ to $t_0 + \tau$).

We can estimate $r_\tau$ from Eq.~(\ref{eq:ODE}) in the short-time limit where the velocity term is insignificant:
\begin{equation}
\ddot{r} - \frac{q}{\tau}\frac{r}{(t + t_0)} = 0.
\label{eq:accel}
\end{equation}

The solution of this equation can be represented in terms of Bessel functions~\cite{Kamke77}, although the complete solution is cumbersome.
In order to avoid dealing with expressions containing special functions we can deduce a simplified estimation of $r_\tau$ based on the following kinematic consideration.
%

After being hit by the front shock wave the particle begins to accelerate (driven by fluid drag), so its trajectory is given by the expression
\begin{equation}
r = r_0 + \frac{a}{2} t^2,
\label{eq:raccel}
\end{equation}
where $a$ is the particle acceleration.
This expression reflects that at $r(t\!=\!0)\!=\!r_0$ and $\dot{r}(t\!=\!0)\!=\!0$ (initially the particle is at rest).
The acceleration term can be estimated by matching Eq.~(\ref{eq:raccel}) with the analytical solution of Eq.~(\ref{eq:accel}) or by directly substituting the expression (\ref{eq:raccel}) into this equation (acceleration being the $\ddot{r}$ term in Eq.~(\ref{eq:accel}) evaluated at $r\!=\!r_0$ and $t\!=\!0$), so $a \simeq q (r_0/t_0)/\tau$.
Then for $t = \tau$, Eq.~(\ref{eq:raccel}) leads to a complete description of the reference position

%
%

\begin{equation}
r_{ref} =  r_0 +   r_0 \frac{q}{2} \left( \frac{\tau}{t_0} \right),
\label{eq:rref}
\end{equation}
which allows us to write the scaling law (\ref{eq:scaleref}) in the following form

\begin{equation}
\frac{r}{r_0} = \left(1 + \frac{q}{2} \frac{\tau}{t_0}\right)\left(\frac{t}{t_0 + \tau} \right)^q.
\label{eq:rmassless}
\end{equation}

Eq.\ (\ref{eq:rmassless}) is the main result of the present study.
We can see that the exponent in this scaling law of particle displacement depends only on the properties of the media (since $q = 2/(5\gamma)$), and is independent of both the properties of the particles and energy of the explosion.
Moreover, since $\gamma > 1$ \cite{Landau87} the particle dispersion is always slower than the ballistic regime (\textit{i.e.}\ $q < 1$).
In general, the dispersion process can be characterized by two limiting cases, depending on the value of the ratio $\tau/t_0$ in Eq.(\ref{eq:rmassless}).
For given characteristics of the explosion (energy $E$), particle and medium properties (Stokes time $\tau$), the ratio $\tau/t_0$ can be associated with the initial position of the particle $r_0$ by introducing the scale
\begin{equation}
r_* = \beta  \left( \frac{E \tau^2}{\rho} \right)^{1/5}.
\label{eq:lanbda}
\end{equation}

For the particles initially located within the sphere $r \le r_*$ (below we refer to this case as the `near field'), we arrive at a simplified form of the scaling law (\ref{eq:rmassless})
\begin{equation}
\frac{r}{r_0} = Q \left(\frac{t}{t_0} \right)^q, \qquad Q = \left(1 + \frac{q}{2} \frac{\tau}{t_0}\right)\left(\frac{t_0}{\tau}\right)^q.
\label{eq:scale1}
\end{equation}
For the opposite case ($r \ge r_*$, the `far field') $Q = 1$~\footnote{see Appendix \ref{sec:nearfar} for derivations}.
We remark that particle properties can influence the value of $Q$ only in the near field region.

At some point the particle motion described by Eqs.~(\ref{eq:scale1}) will be terminated and the particle will come to rest.
The time of this termination corresponds to a disappearance of the driving velocity $\mathbf{V}(\mathbf{r},t)$ in Eq.~(\ref{eq:force}), or a deviation of the function $\mathbf{V}(\mathbf{r},t)$ from the strong explosion model (\ref{eq:vflow}) (\textit{i.e.}\ when the shock wave significantly dissipates).
A simple estimation of this termination point can be deduced from the following arguments.

It is well known that a spherical shock wave loses energy and eventually transforms into a spherical acoustic wave \cite{Landau87,Naugolnykh98}.
This transformation is governed by an interplay between the non-linear and dissipative processes.
As linear acoustic waves cannot generate a persistent flow \cite{Landau87}, it is apparent that a particle cannot be advected any further when this process begins to dominate.
Assuming that the shock wave transformation is mostly due to non-linear effects, and applying the condition $\dot{R} = c$ to Eq.\ (\ref{eq:dRdt}), we can readily deduce a stopping time

\begin{equation}
 t_s = \left(\frac{2 \beta}{5c}\right)^{5/3}\left(\frac{E}{\rho}\right)^{1/3},
\label{eq:ts}
\end{equation}
where $c$ is the speed of sound in the media and all particle dynamics are confined to $t \ll t_s$.

Analogously, one can deduce an estimation for $t_s$ when the shock wave transformation is given by a dissipation process \cite{Landau87, Naugolnykh98}.
Scaling of the dissipation length is given by a diffusion law $\delta(t) \sim (\nu_* t)^{1/2}$, where $\nu_*$ is a well-known aggregated dissipation coefficient determined by viscosity and thermal conductivity (see Ref.\ \onlinecite{Landau87}, \S96).
By equating $R(t)$, Eq.\ (\ref{eq:Rt}), to $\delta(t)$ we arrive at the following expression for the stopping time determined by dissipation
\begin{equation}
 t_s \simeq \left(\frac{E}{\rho}\right)^{2} \left(\frac{1}{\nu_*}\right)^5,
\label{eq:ts2}
\end{equation}
which yields a different $t_s$ value compared to  Eq.\ (\ref{eq:ts}). In the present study we assume the dissipation coefficient $\nu_*$ is relatively small, and hence the stopping time is dominated by non-linear effects described by Eq.\ (\ref{eq:ts}).

Setting $t\!=\!t_s$ in Eqs.~(\ref{eq:scale1}), we can derive the following scaling law for the maximum particle displacement

\begin{equation}
 \frac{r_{max}}{r_0} \propto r^p_0 E^k\tau^h,
\label{eq:scalelaw}
\end{equation}
with values $p=-5/2, k=1/2+2/(15\gamma), h=1-2/(5\gamma)$ for the near field scaling and $p = -1/\gamma,~ k = 1/(3\gamma), ~ h = 0$ for the scaling in the far field~\footnote{Derivations of these exponents can be found in Appendix \ref{sec:exponents}.}.

The scaling of particle displacement with Stokes time $r_{max} \propto \tau^h$ provides insightful information on the effect of particle properties in the system, since for a spherical particle
\begin{equation}
\tau = \frac{d_p^2\rho_p}{18\mu}.\label{eq:tau}
\end{equation}
Here, $d_p$ is the diameter of the particle and $\rho_p \gg \rho$ its density; $\mu$ is the dynamic viscosity of the media.
For instance, with all other parameters being equal, Eq.(\ref{eq:scalelaw}) predicts the particle displacement scales with the media viscosity (in the near field) as ${r_{max}} \propto \mu^{-1+2/(5\gamma)}$.

\section{Numerical Results}

In order to validate our analytical predictions for the scaling laws of Eq. (\ref{eq:scalelaw}), we numerically solve Eq.~(\ref{eq:ODE}) with parameter ranges of $r_0,\, E$ and $\tau$~\footnote{See Appendix \ref{sec:numode} for implementation details.}.
We use the stopping condition Eq.~(\ref{eq:ts}) to calculate a termination point and evaluate the relative particle displacements $r_{max}/r_0$ and then estimate the scaling exponents $p, k, h$ from the log-log plots.
In each set of simulations we change only one parameter keeping all other parameters constant.
The reported parameter values are selected to represent a large range of conditions which cross the near/far field boundary at $r_*$, with enough data points to recover the predicted scaling exponents. Additionally, these values must reside within the time constraint $t_0 \leq t \leq t_s$.
A multitude of parameter values recover the scaling laws, however we plot a single representational value to remove any possible ambiguity in the results.

As a foundation we model explosions in a diatomic gas (air), for which $\gamma = 7/5$ \cite{Landau87} and $q = 2/7$ in Eq.~(\ref{eq:ODE}).
The dynamic viscosity parameter from Eq.\ (\ref{eq:tau}) is assumed to be a constant value of $\mu \simeq 1.983\!\times\!10^{-5}$ Pa/s (air at room temperature), and the particle density represents steel ball bearings $\rho_p = 7874$ kg/m$\mathrm{^3}$ for all reported results.
The results of analytical predictions and numerical simulations are summarized in Table \ref{tbl:scales}.


\begin{table}[tb]
     \caption{\label{tbl:scales} Numerically recovered scaling exponents of physical system parameters against relative particle displacement, Eq.(\ref{eq:scalelaw}), for an explosion in air  ($\gamma = 7/5$).}
\begin{ruledtabular}
\begin{tabular}{llll}
\multicolumn{2}{c}{Near Field ($r \leq r_*$)} & \multicolumn{2}{c}{Far Field ($r \geq r_*$)} \\
Theory & Recovered & Theory & Recovered \\
\hline\noalign{\smallskip}
$p = -5/2$ ($-2.5$) & $p=-2.47$ & $p=-5/7$ ($-0.71$) & $p=-0.72$ \\
$k=~25/42$ ($0.6$) & $k=~~0.58$ & $k=~~5/21$ ~($0.24$) & $k=~~0.23$ \\
$h=~~5/7$ ~~($0.71$) & $h=~~0.71$ & $ h=~~0$ ~~~~~($0$) &  $h=~~0$  \\
\end{tabular}
\end{ruledtabular}
\end{table}

\begin{figure}[tb]
  \centering
  \includegraphics[width=\columnwidth]{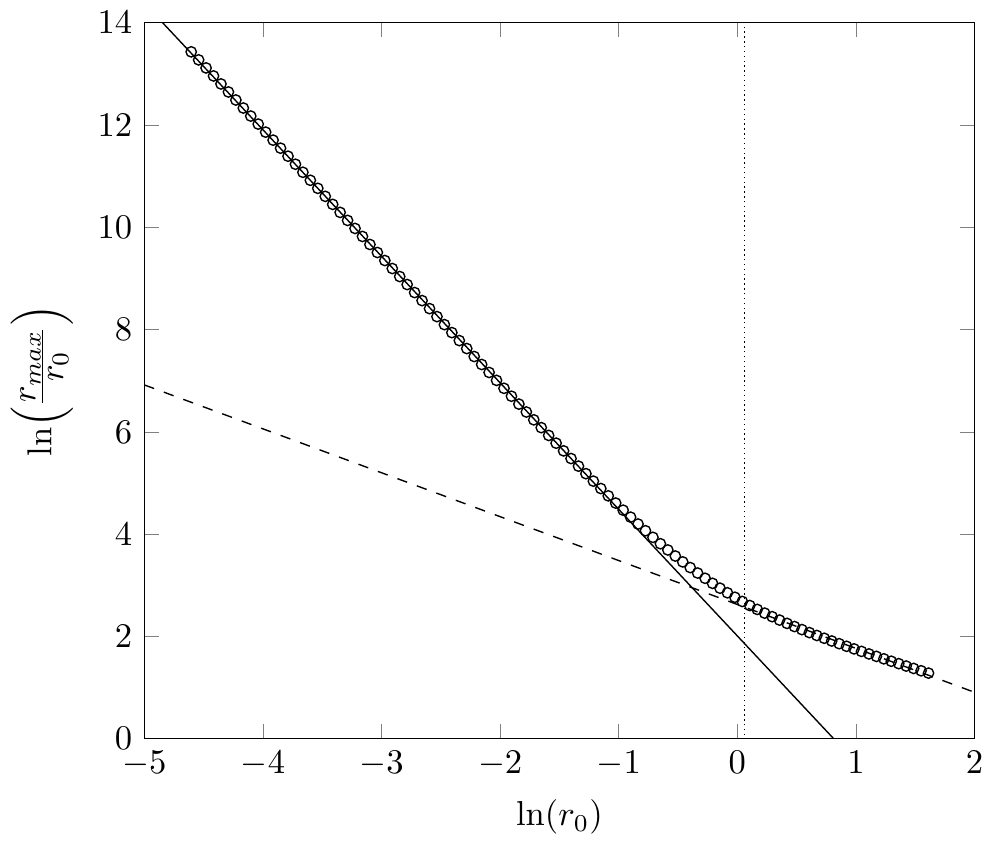}
  \caption{Scaling dependence of the scaled maximum displacement $r_{max}/r_0$ of the numerical solution to Eq.~(\ref{eq:ODE}) against $r_0$ ($\circ$) with $r_*$ indicated as a vertical dotted line. Control parameters for the presented data are $E = 4.52\!\times\!10^{10}$ J/kg and $\tau = 5.52\!\times\!10^{-6}$ s. Recovery of a $-5/2$ dependence (solid line) in the near field and a $-5/7$ dependence (dashed line) in the far field are clearly visible.}\label{fig:zodex0}
\end{figure}

Figure \ref{fig:zodex0} presents a typical output of our simulations.
It depicts the scaling response of a particle's initial position $r_0$ as it is varied between $0.01$ m and $5$ m to the numerical solution of Eq.~(\ref{eq:ODE}).
Two regimes of particle dispersion (near and far field, see Eq.\ (\ref{eq:scalelaw})) are indicated via the solid and dashed lines respectively.
Both results are in good agreement with the theoretical derivations presented in Table~\ref{tbl:scales}.

\begin{figure}[tb]
  \centering
  \includegraphics[width=\columnwidth]{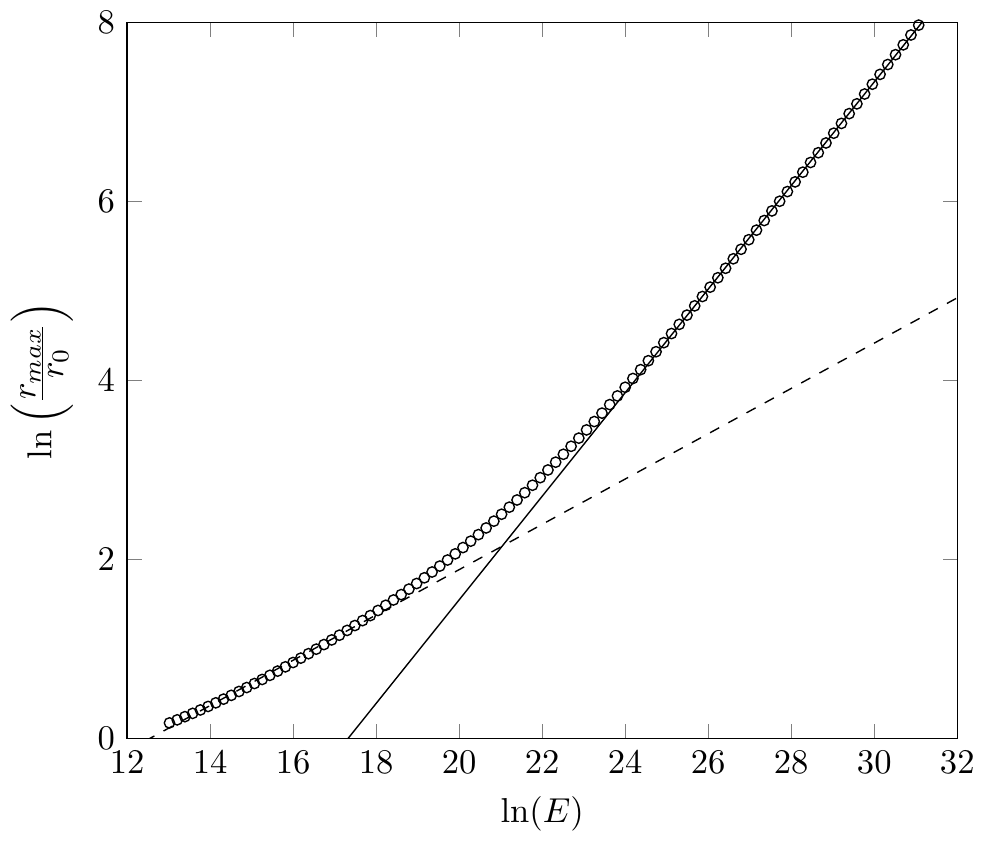}
  \caption{Scaling dependence of $E$ against the scaled maximum displacement $r_{max}/r_0$ of the numerical solution to Eq.~(\ref{eq:ODE}) ($\circ$). Control parameters for the presented data are $r_0 = 0.2$ m and $\tau = 5.52\!\times\!10^{-6}$ s. Recovery of a $25/42$ dependence (solid line) in the near field and a $5/21$ dependence (dashed line) in the far field are clearly visible.}\label{fig:zodeE}
\end{figure}

Energy scaling can be determined in a similar manner, with all other parameters held constant, energy $E$ is adjusted from $452$ kJ/kg to $45.2$ TJ/kg, the result of which can be seen in Figure~\ref{fig:zodeE}.
Here again we recover two scales with good agreement to theory: the near field (solid line) and far field (dashed line) against the numerical results.

\begin{figure}[tb]
  \centering
  \includegraphics[width=\columnwidth]{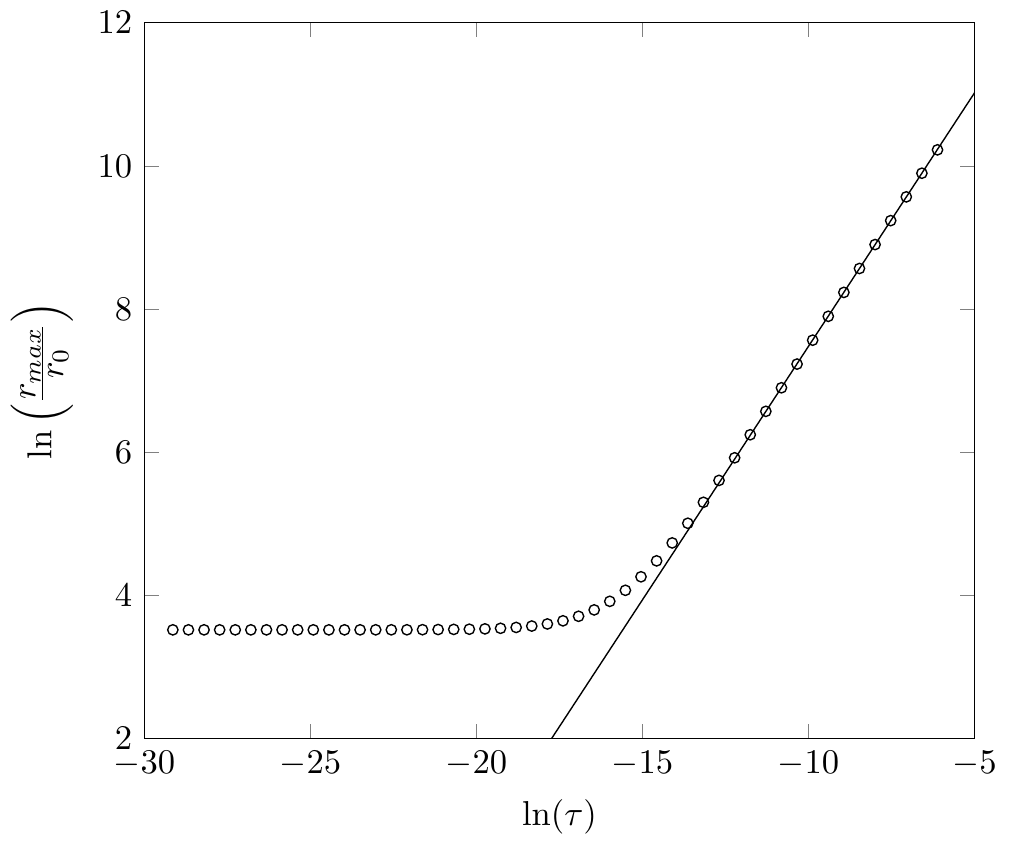}
  \caption{Scaling dependence of $\tau$ against the scaled maximum displacement $r_{max}/r_0$ of the numerical solution to Eq.~(\ref{eq:ODE}) ($\circ$). Control parameters for the presented data are $r_0 = 0.2$ m and $E = 4.52\!\times\!10^{10}$ J/kg. Recovery of a $5/7$ dependence (solid line) in the near field is clearly visible. Displacement is independent of $\tau$ in the far field.}\label{fig:zodetau}
\end{figure}

Finally, we investigate scaling with the Stokes time $\tau$ in Figure~\ref{fig:zodetau}.
As $\tau$ is a function of many parameters (see Eq.~\ref{eq:tau}), we fix all medium variables and alter only the particle diameter (also fixing particle density) over a $1\!\times\!10^{-4}\!-\!10 \;\rm{\mu m}$ range.
Scaling in the near field is recovered with good agreement, and as predicted there is no dependency on $\tau$ in the far field.

To verify the scaling laws derived in Eq.\ (\ref{eq:scalelaw}) are general, we also compare results for noble gasses (\textit{i.e.} when $\gamma = 5/3$ and $\mu \simeq 1.956\!\times\!10^{-5}$ Pa/s for helium at room temperature).
Here, the near field exponents are predicted to be $p=-5/2,\, k=29/50,\, h=19/25$, which recover to $p=-2.5,\, k=0.57,\, h=0.75$ numerically.
In the the far field, $p = -3/5,\, k = 1/5,\,  h = 0$, which also scale as expected recovering $p = -0.62,\, k = 0.19,\,  h = 0$.

\section{Discussion}

The scaling laws established above can provide some predictions that can be validated experimentally.
For instance, it has been observed that there exists a range of particle sizes for which the inertia of the particles combined with the decay of the blast wave allows them to overtake the primary shock front \cite{Zhang01}.
This effect strongly depends on particle size and there exists a threshold (\textit{i.e.} a particle size limit) below which this effect does not occur \cite{Zhang01}.
The proposed framework allows us to formulate a quantitative criteria for this phenomenon.

Consider the dispersion of particles by an explosion in air ($\gamma = 7/5$). The existence of an overtake event directly follows from a comparison of the scaling laws for the position of the shock wave and particle displacement.
The position of the shock wave scales as $t^{2/5}$,  Eq.~(\ref{eq:Rt}), while the particle displacement scales non-uniformly: initially ($t_0 \ll \tau$) it scales as $t^2$ with time, Eq.~(\ref{eq:raccel}), and then slows down to $\propto t^{2/7}$ at the large time limit, Eq.~(\ref{eq:scale1}).
This implies that at the large time limit the particle is always behind the shock wave and can only overtake it during the initial (inertial) stage.
Since particle displacement during the initial stage is given by Eq.~(\ref{eq:raccel}), this leads to the following condition for the particle to be in front of the shock wave:
\begin{equation}
\beta\left(\frac{E t^2}{\rho}\right)^{1/5} = r_0 \left(\frac{t}{t_0}\right)^{2/5} \le r_0 +  (a/2) (t - t_0)^2 , ~~ t \ge t_0.
\label{eq:ovt1}
\end{equation}
By introducing a  new variable $y = (t/t_0)^{1/5}$, this condition can be recast to a non-dimensional form
\begin{equation}
   \Phi(y) =  y^2  - \chi (y^5 -1)^{2}   - 1 \le 0, ~~~ y \geq 1,
\label{eq:ovt2}
\end{equation}
where $\chi =  (q/2)(t_0/\tau) \ll 1$ and $\Phi(1) = 0$.

The function $\Phi(y)$  has a single real root $y_1$ satisfying the condition $y > 1$ with its approximate value $y_1 \approx (1/\chi)^{1/8} \propto (\tau/t_0)^{1/8} \gg 1$.  This root determines the time when the particle catches up and `penetrates through' the decelerating shock wave (since $t\!=\!t_1\!=\!y^5_1 t_0$). This time corresponds to the particle displacement $r_1\!=\!y^{2}_1 r_0 \ge r_0$, after which the particle decelerates and the shock wave overtakes it again.

Similarly, the second time the shock wave overtakes the particle follows from Eqs.~(\ref{eq:Rt}) and (\ref{eq:scale1}):
\begin{equation}
r_1 \left(\frac{t }{t_1}\right)^{2/5} \geq r_1 + r_0  Q  \left(\frac{t - t_1 }{ t_0} \right)^{2/7}, ~~~ t \ge t_1,
\label{eq:ovt3}
\end{equation}
or in a non-dimensional form (substituting the $t_1\!=\!y^5_1 t_0$ and $r_1\!=\!y^{2}_1 r_0$ parameters from the first crossing and $y$ as defined above)
\begin{equation}
     \Gamma(y) = y^{2}  - Q (y^5 - y^5_1 )^{2/7}  - y^2_1  \ge 0, ~~~ y \geq y_1,
\label{eq:ovt4}
\end{equation}
and $\Gamma(y_1) = 0$. The positive real root of $\Gamma(y)$ has an approximate value $y_2 \approx Q^{7/4} \propto (\tau/t_0)^{5/4} \gg 1$.
Both roots ($y_1, y_2$) are dependent on the characteristics of the explosion, and the parameters of medium and particle (via constants $\chi, Q$). It is  evident that the consistency condition, $y_2 \ge y_1$, always holds for sufficiently heavy particles and for particles initially located in proximity to the center of an explosion.

\begin{figure}[tb]
  \centering
  \includegraphics[width=\columnwidth]{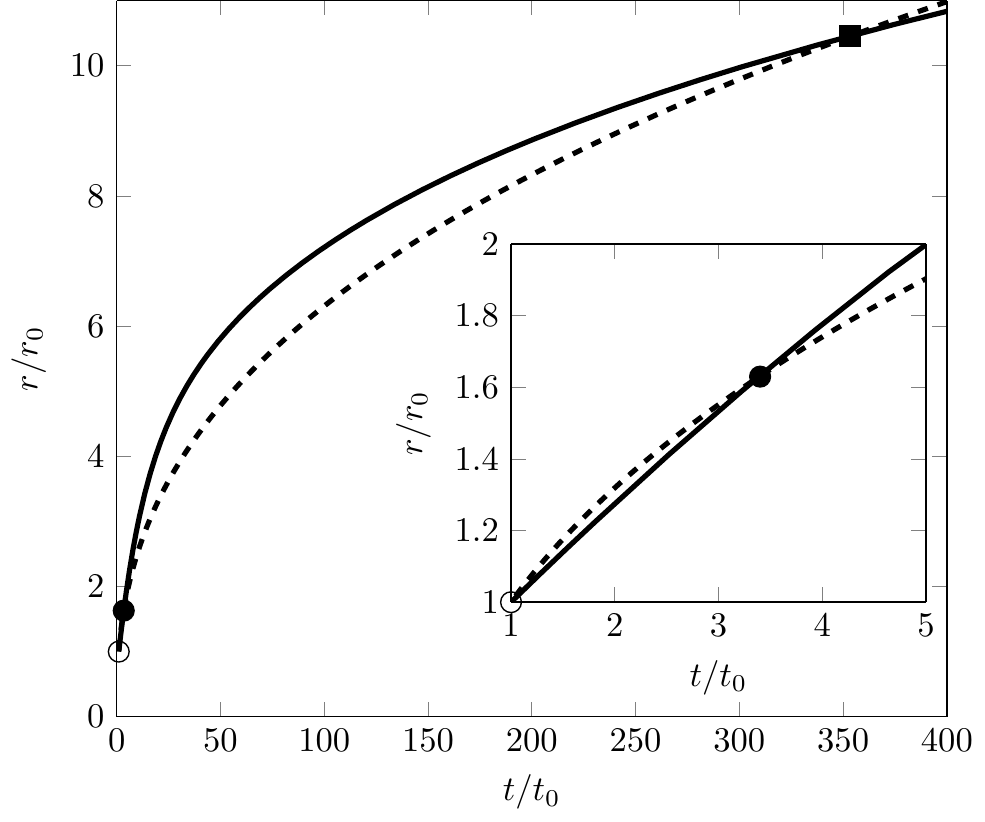}
  \caption{Shock wave trajectory (dashed) and particle position (solid) after an explosion with a yield $E = 0.1$ J/kg. A steel particle with diameter $d_p = 25$ nm is initially at rest at position $r_0 =5$ mm then starts accelerating at time $t_0$ ($\circ$) after the shock passes over. The particle overtakes the shock at $t_1$ ($\bullet$), then at a later time $t_2$ (\mbox{\tiny $\blacksquare$}) it is re-captured and remains behind the shock wave for the rest of the explosive event.}\label{fig:overtake}
\end{figure}

In essence, we have identified three consecutive events where the particle and shock wave cross (one possible set of parameters which observes this phenomena is presented in Figure \ref{fig:overtake}).
At $t\!=\!t_0$ ($r\!=\!r_0$), the shock wave initially reaches a particle which is at rest.
Driven by inertia, the particle overtakes the decelerating shock wave at $t\!=\!t_1 > t_0$ ($r\!=\!r_1$).
Finally, at $t\!=\!t_2 > t_1$ ($r\! =\! r_2$) the shock wave again catches up to the decelerating particle and overtakes it.
The zeroth-order estimates of $t_1$ and $t_2$ derived from Eq.\ (\ref{eq:ovt2}) and (\ref{eq:ovt4}) respectively obtain an acceptable agreement with the numerical results of Figure \ref{fig:overtake}: $t_1 \approx 2\!\times\!10^{-8} \;\mathrm{(est)} = 9\!\times\!10^{-9}\; \mathrm{(numeric)}$ and $t_2 \approx 2.5\!\times\!10^{-7} \;\mathrm{(est)} = 9.5\!\times\!10^{-7} \;\mathrm{(numeric)}$.

An experimental validation of the `wave-particle' overtake phenomena is a challenging undertaking which requires precise (and simultaneous) measurements of the positions of shock waves and tracer particles; as such there are few publications on this subject.
To the best of our knowledge there is only a single experimental study, Ref.~\onlinecite{Zhang01}, in which the `wave-particle' overtake phenomena has been observed and reported.
The positions of the particles and shock in this study have been detected by means of two $150$ kV flash X-ray pulsers and six piezoelectric pressure transducers respectively.
Our interpretation of these phenomena, presented in this paper, stems from a simple kinematic analysis of the scaling laws for particle displacement, is in qualitative agreement with the experimental observations reported in Ref.~\onlinecite{Zhang01}.

Another interesting effect associated with an explosive energy release in a particle system is the formation of a residual sparseness (cavity) in an initially uniform particle distribution (\textit{e.g.}\ dust) after the particles have been displaced by the shock wave from their initial position.
We can easily estimate the scale of this cavity by invoking the analytical framework presented above.
In fact, the edge of the cavity is formed from the far field particles, for which  $t_0 \gg \tau$.
By setting $t\!=\!t_s$ and $r \sim  r_0 \sim r_{cav}$ in Eq.\ (\ref{eq:scale1}) we can derive an estimate

\begin{equation}
 r_{cav} \simeq \kappa \left(\frac{E }{\rho c^2}\right)^{1/3},
\label{eq:cav1}
\end{equation}
where $\kappa = (4 \beta^5/25)^{1/3} \approx 0.09 $, which can also be established based on dimensional arguments.
This expression provides a characteristic scale of the density of particle distribution associated with the residual sparseness caused by an explosion (for a visual example of such a cavity, see images in Ref.\ \onlinecite{Antal08} and Figure 1 of Ref.\ \onlinecite{Harris15}).

\section{Conclusions}

In summary, we have presented scaling laws for particle displacement caused by a rapid (and localized) energy release (explosion).
These scaling laws account for particle properties (mass, diameter, density), properties of the medium a particle is advected through (viscosity, density, speed of sound) and the energy of the explosion dissipated through the shock front.
We demonstrated that by employing a conceptual model of particle displacement that includes the strong explosion model and simplified particle kinematics, the exponents of these scaling laws can be derived analytically, which are in good agreement with numerical simulations.

This framework has been constructed in a manner such that more realistic descriptions (different models of explosion, non-Stokes drag, multi-phase transformations, \textit{etc}) can be implemented to ascertain improved estimations of each scaling law outlined here.
For instance, taking inertial effects of the fluid flow into account leads to an equation of motion with a different form of drag~\cite{Landau87}
\begin{equation}
\dot{\textbf{r}} = \textbf{v}, \quad \dot{\textbf{v}} = \frac{1}{\lambda}(\textbf{V}  - \textbf{v})^2,
\end{equation}
where the parameter $\lambda$ has a dimension of length and provides an aggregated characteristic of particles and media (size, density, viscosity), similar to Eq.\ (\ref{eq:tau}).
It is evident from this equation that the same scaling laws for particle displacement exist in the far field, \emph{viz}., with the exponents given in Table \ref{tbl:scales}.
This and further additions will be elaborated upon in our future work.

We anticipate that the results presented in this study will be useful in evaluating high fidelity models of particle transport and interpreting experimental observations of explosion phenomena.

\appendix
\section{Near and Far Field Limits}\label{sec:nearfar}

Eq.\ (\ref{eq:rmassless}) is characterised by two limiting conditions, dependent on the ratio $\tau/t_0$.

We initially investigate the `far field', where $\tau/t_0 \ll 1$ and $r \geq r_*$ by taking the limit
\begin{equation}
\lim_{\tau/t_0 \to 0} \frac{r}{r_0} = (1+0)\left(\frac{t}{t_0}\right)^q
\end{equation}
and defines the simple scaling law
\begin{equation}
    \frac{r}{r_0} = Q \left(\frac{t}{t_0} \right)^q, \quad Q = 1.
\end{equation}

The `near field' conditions are obtained via the condition $\tau/t_0 \gg 1$ ($r < r_*$). This limit is given by
\begin{equation}
\lim_{\tau/t_0 \to \infty} \frac{r}{r_0} = \left(1 + \frac{q}{2} \frac{\tau}{t_0}\right)\left(\frac{t}{\tau}\right)^q
\end{equation}
and defines the near field scaling law as
\begin{equation}
    \frac{r}{r_0} = Q \left(\frac{t}{t_0} \right)^q, \quad Q = \left(1 + \frac{q}{2} \frac{\tau}{t_0}\right)\left(\frac{t_0}{\tau}\right)^q.
\end{equation}

\section{Scaling Law of Maximum Particle Displacement}\label{sec:exponents}

To derive the scaling law for maximum particle displacement in Eq.\ (\ref{eq:scalelaw}) we set $t=t_s$ in Eqs.\ (\ref{eq:scale1}) such that
\begin{align}
\frac{r_{max}}{r_0} &= Q\left(\frac{t_s}{t_0}\right)^q \\
&= Q\left[\frac{\left(\frac{2 \beta}{5}\right)^{5/3}\left(\frac{E}{\rho c^5}\right)^{1/3}}{\left(\frac{r_0}{\beta}\right)^{5/2}\left(\frac{\rho}{E}\right)^{1/2}}\right]^q\\
&= Q\left[\left(\frac{2}{5}\right)^{5/3}\beta^{25/6}\left(\frac{E}{\rho}\right)^{5/6}c^{-5/3}r_0^{-5/2}\right]^q.
\end{align}

The general case, where $q=2/(5\gamma)$, we arrive at
\begin{equation}
\frac{r_{max}}{r_0} = Q\left[\left(\frac{2}{5}\right)^{\frac{2}{3\gamma}}\beta^{\frac{5}{3\gamma}}\left(\frac{E}{\rho}\right)^{\frac{1}{3\gamma}}c^{-\frac{2}{3\gamma}}r_0^{-\frac{1}{\gamma}}\right].\label{eq:rmaxfar}
\end{equation}

For the far field, $Q = 1$ and we recover the scaling conditions
\begin{equation}
\frac{r_{max}}{r_0} \propto r_0^{-\frac{1}{\gamma}}E^{\frac{1}{3\gamma}}\tau^0.
\end{equation}

The near field requires further treatment as $Q$ in this regime is non-scalar.
\begin{align}
Q &= \left[1 + \frac{q}{2} \frac{\tau}{\left(\frac{r_0}{\beta}\right)^{5/2}\left(\frac{\rho}{E}\right)^{1/2}}\right]\left[\frac{\left(\frac{r_0}{\beta}\right)^{5/2}\left(\frac{\rho}{E}\right)^{1/2}}{\tau}\right]^q \\
&= \left(\frac{r_0}{\beta}\right)^{5q/2}\left(\frac{\rho}{E}\right)^{q/2}\tau^{-q}\nonumber\\
&\qquad\quad +\frac{q}{2}\left(\frac{r_0}{\beta}\right)^{-5/2+5q/2}\left(\frac{\rho}{E}\right)^{-1/2+q/2}\tau^{1-q}\\
&= \left(\frac{r_0}{\beta}\right)^{1/\gamma}\left(\frac{\rho}{E}\right)^{1/5\gamma}\tau^{-2/5\gamma}\nonumber\\
&\quad +\frac{1}{5\gamma}\left(\frac{r_0}{\beta}\right)^{-5/2+1/\gamma}\left(\frac{\rho}{E}\right)^{-1/2+1/5\gamma}\tau^{1-2/5\gamma}\label{eq:qgam}
\end{align}

Substituting Eq.\ (\ref{eq:qgam}) into Eq.\ (\ref{eq:rmaxfar}), we find
\begin{widetext}
\begin{align}
\frac{r_{max}}{r_0} &= \nonumber\\ &\left[\left(\frac{r_0}{\beta}\right)^{\frac{1}{\gamma}}\left(\frac{\rho}{E}\right)^{\frac{1}{5\gamma}}\tau^{-\frac{2}{5\gamma}}+\frac{1}{5\gamma}\left(\frac{r_0}{\beta}\right)^{-\frac{5}{2}+\frac{1}{\gamma}}\left(\frac{\rho}{E}\right)^{-\frac{1}{2}+\frac{1}{5\gamma}}\tau^{1-\frac{2}{5\gamma}}\right]\left[\left(\frac{2}{5}\right)^{\frac{2}{3\gamma}}\beta^{\frac{5}{3\gamma}}\left(\frac{E}{\rho}\right)^{\frac{1}{3\gamma}}c^{-\frac{2}{3\gamma}}r_0^{-\frac{1}{\gamma}}\right]\\
&= \left(\frac{2\beta}{5c}\right)^{\frac{2}{3\gamma}}\left(\frac{\rho}{E}\right)^{-\frac{2}{15\gamma}}\tau^{-\frac{2}{5\gamma}}+\frac{1}{5\gamma}\left(\frac{2}{5}\right)^{\frac{2}{3\gamma}}\beta^{\frac{5}{2}+\frac{2}{3\gamma}}c^{-\frac{2}{3\gamma}}r_0^{-\frac{5}{2}}\tau^{1-\frac{2}{5\gamma}}\left(\frac{\rho}{E}\right)^{-\frac{1}{2}-\frac{2}{15\gamma}}.
\end{align}
\end{widetext}

This result reveals two scales in the near field, with a critical point at
\begin{equation}
N_* = \frac{5\gamma}{\tau}\left(\frac{r_0}{\beta}\right)^{5/2}\left(\frac{\rho}{E}\right)^{1/2} = \frac{5\gamma t_0}{\tau},
\end{equation}

representing the position at which the acceleration term dominates the initial position term of Eqs.\ 10 and 11 in the main text.
$N_*$ is inversely proportional to $\tau/t_0$ and as such $N_* \gg 1$ represents very-small times, with $N_* \ll 1$ being the primary term.

We can now recover the scaling conditions
\begin{align}
\left.\frac{r_{max}}{r_0}\right|_{N_* \gg 1} &\propto r_0^{0}E^{\frac{2}{15\gamma}}\tau^{-\frac{2}{5\gamma}},\\
\left.\frac{r_{max}}{r_0}\right|_{N_* \ll 1} &\propto r_0^{-\frac{5}{2}}E^{\frac{1}{2}+\frac{2}{15\gamma}}\tau^{1-\frac{2}{5\gamma}}.
\end{align}

$N_* \gg 1$ exists only in the limit $t\to t_0$ and can therefore be dropped from consideration.

Finally, as an example of scaling conditions, we can use $\gamma = 7/5$ (air) to obtain the values for the near field quoted in Table \ref{tbl:scales} of the main text: $p=-5/2,\; k=25/42,\; h=5/7$.

\section{Numerical Treatment of the Eq.\ (\ref{eq:ODE}) ODE}\label{sec:numode}

The second order ODE of Eq.\ (\ref{eq:ODE}) was decoupled to a set of first order ODEs.
Let $y_1=y$ and $y_2=\dot{y}$, giving the first order system

\begin{align}
\dot{y_1} &= y_2 \\
\dot{y_2} &= -\frac{1}{\tau}y_2+\frac{q}{\tau}\frac{y_1}{(t+t_0)}.
\end{align}

This system was then solved in \texttt{MATLAB} using the \texttt{ode15s} stiff solver, which implements a variable order (variable step size) method based on finite difference formulas.

\bibliography{shock}

\end{document}